\documentclass[%
 reprint,
 superscriptaddress,
 twocolumn,
 amsmath,
 amssymb,
 longbibliography,
 aps,
 nofootinbib,
 pra
]{revtex4-2}

\usepackage{graphicx}
\usepackage{dcolumn}
\usepackage{bm}
\usepackage{hyperref}

\usepackage{epsfig}
\usepackage[normalem]{ulem}
\usepackage{siunitx}
\usepackage{verbatim}

\begin{document}

\preprint{}

\title{Photonic heat amplifiers based on a disordered semiconductor}

\author{Matteo Pioldi}
 \email{matteo.pioldi@sns.it}
\author{Giorgio De Simoni}%
\author{Alessandro Braggio}
\author{Francesco Giazotto}
 \email{francesco.giazotto@sns.it}
\affiliation{%
 NEST, Istituto Nanoscienze-CNR and Scuola Normale Superiore, I-56127 Pisa, Italy
}%


\begin{abstract}
A photonic heat amplifier (PHA) designed for cryogenic operations is introduced and analyzed. This device comprises two variable-range-hopping reservoirs connected by lossless lines, which allow them to exchange heat through photonic modes. This configuration enables negative differential thermal conductance (NDTC), which can be harnessed to amplify thermal signals. To achieve this, one reservoir is maintained at a high temperature, serving as the source terminal of a thermal transistor. Concurrently, in the other one, we establish tunnel contacts to metallic reservoirs, which function as the gate and drain terminals. With this arrangement, it is possible to control the heat flux exchange between the source and drain by adjusting the gate temperature.
We present two different parameter choices that yield different performances: the first emphasizes modulating the source-drain heat current, while the second focuses on the modulation of the colder temperature variable range hopping reservoir. Lastly, we present a potential design variation in which all electronic reservoirs are thermally connected through only photonic modes, allowing interactions between distant elements. The proposal of the PHA addresses the lack of thermal transistors and amplifiers in the mK range while being compatible with the rich toolbox of circuit quantum electrodynamics. It can be adapted to various applications, including sensing and developing thermal circuits and control devices at sub-Kelvin temperatures, which are relevant to quantum technologies. 
\end{abstract}

\maketitle

\section{\label{sec:intro}Introduction}

The management of heat and temperature is a crucial issue in quantum technologies. Addressing this challenge opens up experimental opportunities and advances to the state-of-the-art
\cite{fornieri_2017}. To achieve control over thermal properties comparable to electrical ones, it is essential to develop heat current and temperature amplifiers. The electrical transistor, vital to the success of conventional electronics, demonstrates amplifying behavior. Thus, creating a thermal equivalent operating in extremely low-temperature regimes would signify a breakthrough in heat management of quantum nanotechnologies. Developing a thermal transistor and amplifier could enable better control of heating and dissipation, which can be beneficial to the functioning of nanoscale devices, especially for semiconductor-based ones, in which less efficient heat relaxation is observed \cite{xie_2023, eduardolee2024heating}. Furthermore, like its electrical analog, a heat transistor could be featured in thermal logic circuits \cite{li2007logic} and passive elements for autonomous temperature regulation on a chip.

Proposals and experimental studies on heat transistors have been conducted in various contexts, including room-temperature devices \cite{swoboda_2021,castelli_2023} and nanoscale phononic heat transport \cite{xie_2023}. However, there remains a gap to be addressed for devices operating within the framework of superconducting quantum technologies \cite{saira2007heat}, where a few tenths of \unit{\milli \kelvin} are typically considered for applications such as computing \cite{krinner_2019,somoroff_2023} or high-sensitivity measurements \cite{romanenko_2020}. At such low temperatures, phononic degrees of freedom quickly become an ineffective heat transport medium, while quasiparticle diffusion \cite{giazotto_2006,fornieri_2017} and photonic radiation mechanisms begin to prevail \cite{meschke2006single,ojanen2008mesoscopic,partanen_2016,ronzani2018tunable,subero2023resistive,iorio2021photonic,pekola2021colloquium,marchegiani_diode_2021,senior2020heat,gubaydullin2022photonic}. 

This article discusses a heat transistor that amplifies thermal signals through a combination of electronic and photonic channels, the latter being essential, leading to the \textit{photonic heat amplifier} (PHA) design. As we shall see, this design can also be adapted to work with only photonic heat transport channels between its terminals. Using photons enables efficient heat transport in a sub-Kelvin environment. We will examine the performance of the proposed design as a heat current or temperature modulation amplifier by utilizing negative differential thermal conductance (NDTC), as initially suggested in Ref. \cite{casati_2006}. 
Even if there are other thermal amplifiers based on inelastic effects \cite{jiang_2015, wang_2022}, we believe that using NDTC will enable operating at extremely low temperatures, but still with heat flows of the order of hundreds \unit{\femto\watt}, thereby providing sufficiently strong heat current control. 
In our proposal, NDTC occurs due to the sensitivity of the thermal transmission coefficients to the reservoirs' impedance matching \cite{pascal_2011}, which may vary with temperature, as we will see. Specifically, NDTC arises from a photonic thermal contact between two reservoirs made of a disordered semiconductor with variable range hopping (VRH) \cite{efros_1975}, where the efficiency of heat transport $\tau$ counterintuitively increases as the temperature difference between the terminals decreases. This behavior allows for NDTC and thermal amplification when effectively integrated into a properly designed device, as discussed later.

\begin{figure*}[t!]
    \centering
    \includegraphics[width=0.97\linewidth]{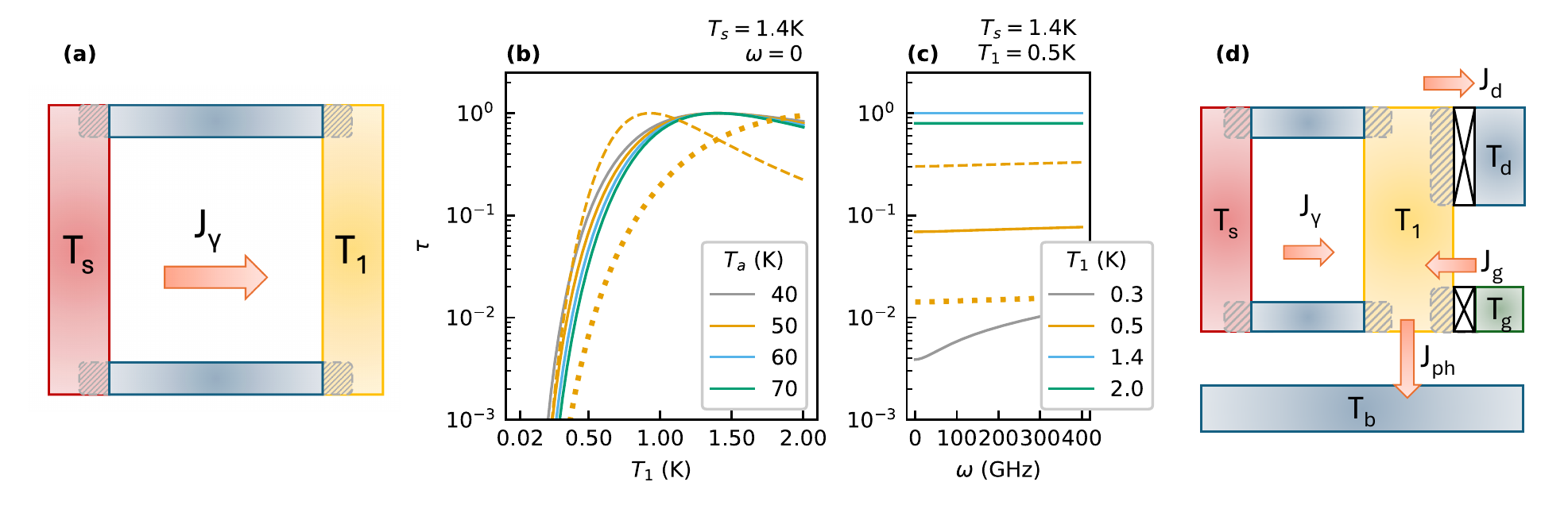}
    \caption{\textbf{(a)} A photonic thermal link connects two VRH reservoirs. A hotter ``source" reservoir at temperature $T_s$ (shown in red) exchanges heat with a colder reservoir at temperature $T_1$ through photonic modes that travel across lossless wires (indicated in blue), contributing to a heat flux $J_{\gamma}$. The etched areas denote regions that are heavily doped to create low-resistance Ohmic contacts. 
    \textbf{(b)} The thermal transmission coefficient $\tau$, as defined in Sec. \ref{sec:ndtc}, illustrates how efficiently heat
    is transmitted from one reservoir to another. The plot displays $\tau$ for different values of $T_a$ while keeping $T_s=\qty{1.4}{\kelvin}$ and $\omega=0$ constant as $T_1$ is adjusted. For $T_a = \qty{50}{\kelvin}$ both the perfectly-matched resistivity $\chi=1$ (continuous line) and two mismatched configurations
    ($\chi=5$, dashed line, and $\chi=0.2$, dotted line) are shown, while the other curves are drawn for $\chi=1$. An interval for which $T_1>T_s$ is depicted for the sake of completeness and clarity, highlighting how, for $\chi=1$, the perfect matching precisely happens at $T_s=T_1$.
    \textbf{(c)} The weak dependence with $\omega$ of the thermal transmission coefficient $\tau$ for $T_a=\qty{50}{\kelvin}$, $T_s=\qty{1.4}{\kelvin}$ for different values of $T_1$. The $T_1=\qty{2}{\kelvin}>T_s$ is included to show what happens after that a perfect matching for $\chi=1$ is achieved at $T_1=T_s$. For $T_1=\qty{0.5}{\kelvin}$, both the $\chi=1$ case (continuous line) and two other configurations ($\chi=5$, dashed line, and $\chi=0.2$, dotted line) are shown, while the other curves are drawn for $\chi=1$.
    \textbf{(d)} The proposed device features a central island at temperature $T_1$ that transfers heat currents (defined in Eqs. \ref{eq:Jgamma},\ref{eq:Jt}-\ref{eq:Jph}) to various terminals. The tunnel contacts to the drain and gate are positioned at heavily doped regions of the yellow central island, highlighted by the same gray etched pattern from panel (a). Each arrow indicates the positive direction of the represented heat flux. The phonon bath is maintained at temperature $T_b$, the gate (in purple) is at $T_g$, and the drain (in blue) is at $T_d$, always taken equal to $T_b$.}
    \label{fig:0-intro}
\end{figure*}
This paper is organized as follows: Sec. \ref{sec:ndtc} explains how photonic NDTC is achieved; Sec. \ref{sec:device} provides a comprehensive overview of the PHA, including a discussion of the relevant regimes and the impact of the device parameters; Sec. \ref{sec:th-ampli} details the configurations of the current modulation amplifier (CMA) and the temperature modulation amplifier (TMA), also showing a potential alternative design that is essentially a fully photonic heat transistor; finally, Sec. \ref{sec:concl} concludes by summarizing the potential of the PHA and offering insights into future developments.
 
\section{\label{sec:ndtc}Photonic NDTC}

Unlike other proposals based on the interaction of reservoirs with two- and three-level systems 
\cite{majland_2020}, here NDTC is obtained by relying on photonic transport and temperature-dependent material properties. 
Photonic heat transport is the primary pathway for heat exchange between electronic reservoirs located below \qty{1}{\kelvin}, linked by lossless superconducting lossless lines maintained well below the critical temperature of the superconductor \cite{schmidt2004photon}, a typical regime for conventional dilution cryostats used in superconducting nanotechnologies. 
In fact, placing two reservoirs in thermal contact via two superconducting lines, kept colder than the critical temperature, inhibits any exchange of quasiparticles, and heat flow is mainly determined by electromagnetic fluctuations in the lines (photons) \cite{marchegiani_diode_2021}.
 A system in which heat transport occurs through photonic modes has been proposed to achieve NDTC \cite{marchegiani_2023}. The impedance matching of the involved reservoirs significantly influences the heat exchange efficiency of such a mechanism. 

To illustrate this concept, we adopt the simplified lumped element approximation and examine heat transport along a lossless line between two electronic reservoirs with impedances $Z_s$ and $Z_1$ (see Fig. ~\ref{fig:0-intro}(a) for a diagram of such a system). In our proposal, we can neglect any capacitive and inductive components of the reservoirs. We assume purely dissipative impedances, at least for the regime of energies more relevant to heat transport. To analyze the origin of the NDTC, we assume that the reservoirs are operating at temperatures $T_s$ and $T_1$, respectively. It is crucial to mention that, generally, the electronic temperatures may influence the reservoir impedances. 
The lossless lines (in blue) are assumed to be thermalized at bath temperature, and the arrow indicates the photonic heat current $J_{\gamma}$, which is considered positive when $T_s > T_1$ (flowing in the direction of the arrow). 

The photonic heat current $J_{\gamma}(T_s,T_1)$ depends on the thermal transmission coefficient $\tau(\omega,T_s,T_1)$ \cite{pascal_2011} as
\begin{eqnarray}
    \tau(\omega,T_s,T_1) = 4\frac{\operatorname{Re}[Z_s(\omega,T_s)] \operatorname{Re}[Z_1(\omega,T_1)]}{|Z_s(\omega,T_s) + Z_1(\omega,T_1)|^2}.
    \label{eq:tau}
\end{eqnarray}
For two identical reservoirs whose impedance is monotonous in temperature, $\tau$ should tend to 1 as $T_s$ and $T_1$ approach each other, due to the improvement in impedance matching.
This is precisely illustrated in Fig. \ref{fig:0-intro}(b), where $\tau$ is evaluated for $\omega=0$. Fig. \ref{fig:0-intro}(c) clarifies that, for our parameter choice, in the operating range $T_1 \gtrsim \qty{0.5}{\kelvin}$, the frequency dependence is practically negligible.
Exploiting the knowledge of the expression for $\tau$, the expression for $J_{\gamma}$ becomes \cite{pascal_2011}:
\begin{eqnarray}
    J_{\gamma}(T_s,T_1) = \int_0^\infty \frac{\hbar \omega}{2 \pi} \tau(\omega, T_s,T_1) [n_s(\omega)-n_1(\omega)],
\label{eq:Jgamma}
\end{eqnarray}
where $n_i(\omega)=(\exp[\hbar \omega/(k_B T_i)]-1)^{-1}$.
This formula lets us appreciate that if the temperature $T_s$ is held constant and $T_1$ is increased from a low base value, two competing effects impact the heat current $J_{\gamma}$ as the thermal gradient $(T_s - T_1)$ decreases. On the one hand, the gradient reduction should decrease $J_{\gamma}$; on the other hand, the improved efficiency $\tau$, resulting from the better impedance matching between $Z_s(T_s)$ and $Z_1(T_1)$, tends to increase it. 
When the latter effect dominates, NDTC is achieved because the heat current increases even with the declining thermal gradient. 
The PHA will be obtained by integrating a photonic thermal channel, thereby achieving NDTC in an aptly designed three-terminal device, as shown in Fig. \ref{fig:0-intro}(c).

\subsection{NDTC with VRH}

\color{black}
Reservoirs with a strong temperature dependence of the impedance are suitable for achieving NDTC behavior. Material systems showing the VRH phenomenology may meet these requirements. 
The DC resistance $R_i$ of a reservoir made of this type of materials can be expressed as \cite{efros_1975}:
\begin{eqnarray}
    R_{i,DC}(T) = R_{0,i} \exp[\sqrt{T_a/T}],
    \label{eq:ai-rho}
\end{eqnarray}
where $T_a$ is a constant in the $1 - 100 \unit{\kelvin}$ range and the $i = s,1$ subscript identifies the reservoir. $R_{0,i}$ is a constant depending on the material properties and the geometry of the reservoir. For example, this behavior is typically exhibited by neutron-transmutation-doped (NTD) germanium \cite{beeman_1990, beeman_1996, mccammon_2005}. It is important to note that electrical fluctuation in the proposed devices, in the operating regime we will consider,  will be very low such that the non-Ohmic behavior (see, for example, Ref. \cite{mccammon_2005}) does not affect our predictions. However,  the reservoir impedance could also present a frequency-dependent part at finite frequencies. For our example, we will refer to the theoretical work of Refs. \cite{shlovskii1981AC, bottger1979AC} and the experimental observation of Ref. \cite{jang1990infrared} regarding NTD germanium. Thus, the resistivity $\rho_i$ of NTD germanium, in a regime where $\hbar \omega, k_B T_i \ll k_B T_a$, can be estimated with:
\begin{widetext}
    \begin{eqnarray}
    \rho_i(\omega, T_i) = [\sigma_{DC}(T_i) + \sigma_{AC}(\omega,T_i)]^{-1} = \frac{1}{\sigma_{0,DC}\exp[-\sqrt{T_a/T_i}] + A \omega (1-\exp[-\hbar \omega / k_B T_i])},
    \label{eq:rho-omega}
\end{eqnarray}
\end{widetext}
where $\sigma_{DC,AC}$ are the DC and AC conductivity, respectively. Our experimental-informed estimates of these parameters are $\sigma_{DC} = \qty{1.5}{(\ohm \micro \metre)^{-1}}$ \cite{pasca2004thesis} and $A = \qty{0.05}{(\ohm \metre \giga \hertz)^{-1}}$ \cite{jang1990infrared} for $T_a=\qty{50}{\kelvin}$. Since $T_a$, $A$ and $\sigma_{DC}$ depend on the material doping, they are interrelated. When varying $T_a$, the other parameters will also be adjusted according to the theoretical models to which we refer. 
The high values of $T_a > \qty{10}{\kelvin}$ define an energy scale for the resistance behavior which sets the working point of the device and is at the same time much higher than the operating temperatures $T_s$, $T_1$ and the related frequencies $\omega$, allowing a linear model for the description of heat transport and thermal fluctuations.   
The transmission coefficient $\tau$ between two VRH reservoirs with the same $T_a$ residing at temperatures $T_s$ and $T_1$ is:
\begin{eqnarray}
    \tau(\omega,\chi,T_a;T_s,T_1) = & 4\frac{\chi \rho_s(\omega, T_s) \rho_1(\omega, T_1)}{(\chi \rho_s(\omega, T_s) + \rho_1(\omega, T_1))^2},
    \label{eq:ai-tau}
\end{eqnarray}
where $\chi$ is a factor inserted to account for geometrical differences between the terminals, and it can be even redefined to include cases with different values of $T_a$ between the two reservoirs or other corrections.

This expression assumes that the contacts between the lossless lines and the reservoirs can be neglected with respect to the intrinsic semiconductor resistances. Indeed, a low-resistance contact with such reservoirs can be achieved by heavily doping the region corresponding to the contact itself, making the material locally metallic. This technique is commonly used in semiconductor materials to create Ohmic contacts \cite{beeman_1990}.
The heavily doped metallic regions in our images will present an etched pattern, as shown in Fig. \ref{fig:0-intro}(a) near the contacts with the superconductor. We assume that the electrons in these heavily doped regions thermalize with the other electrons in the reservoirs on a timescale much shorter than the device's operating timescale. 
 
The dependence of $\tau$ on $T_1$ is plotted in Fig. \ref{fig:0-intro}(b), with $T_s$ kept fixed at $\qty{1.4}{\kelvin}$, $\omega=0$ and $\chi = 1$. This behavior plotted for $\omega=0$ is indeed representative, for high enough $T_1$, for the efficiency of the transmission at all frequencies. Furthermore, Fig. \ref{fig:0-intro}(c) shows that for $T_1 \gtrsim \qty{0.5}{\kelvin}$, the frequency dependence is almost negligible for the typical operating point of our device. At the same time, it contributes fundamentally to the low-temperature behavior because in such case the DC conductivity becomes negligible with respect to the AC conductivity. However, for the typical operating point of our device this frequency dependence can be safely neglected. The frequency range covers the values where the factor $n(\omega,T_s)-n(\omega,T_1)$ is not negligible for the temperatures discussed in this article. In the plots, we consider values of $T_1$ related to such ranges.

In Fig. \ref{fig:0-intro}(b,c), we can observe that, for values of $T_1$ close to the base, $\tau$ tends to 0. This is because $Z_s(\omega, T_s=\qty{1.4}{\kelvin})$ and $Z_1(\omega, T_1)$ are very mismatched, suppressing heat flow. Instead, for $T_1=T_s=\qty{1.4}{\kelvin}$, when $\chi=1$, the nominal maximum value $\tau=1$ is achieved, thanks to the perfect matching of the two impedances. For completeness, we added a curve with a value of $T_1$ higher than $T_s$: in this case, the increasing difference between $Z_s$ and $Z_1$ lowers the value of $\tau$, as can be seen in the curve $T_1 = \qty{2}{\kelvin}$. $T_1 > T_s$ will not be considered in the following, keeping the source temperature (see later) larger than the other, \textit{i.e.} $T_s > T_1$.
We point out again that, as the temperature increases, the DC resistivity becomes smaller and $\rho$ varies less with frequency in the range we are considering, in line with the expectations of Ref. \cite{shlovskii2024review}. In addition, two configurations of mismatched resistivity, $\chi=5$ (dashed line) and $\chi=0.2$ (dotted line) are considered for $T_a=\qty{50}{\kelvin}$. We need to remember that in NTD germanium, resistivity decreases with increasing temperature. Consequently, when $\chi=5$, a better match is achieved for some values $T_1<T_s$, while for $\chi=0.2$, we would need higher values of $T_1$ for a good match. We will focus on $\chi=5$ for the remainder of the article, to harness a higher transmission coefficient at low temperatures, where a better performance of the photonic contribution is expected.

\begin{figure}[t!]
    \centering
    \includegraphics[width=\linewidth]{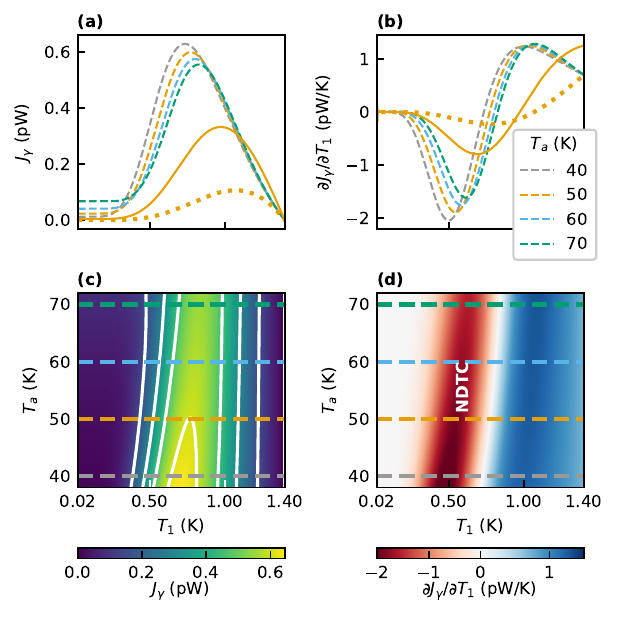}
    \caption{In all the plots of this figure, quantities are evaluated for $T_s=\qty{1.4}{\kelvin}$ and $\chi=5$. \textbf{(a-b)} The dependence of the photonic heat current $J_{\gamma}$ (see Eq. \ref{eq:Jgamma}) and the differential thermal conductance $g_{\gamma}$ (see Eq. \ref{eq:g-gamma}) on $T_1$ for different values of $T_a$. For $T_a=\qty{50}{\kelvin}$, $\chi=1$ (continuous line) and $\chi=0.2$, dotted line, are also shown. \textbf{(c)} A color map of the intensity of the heat current between the reservoirs as a function of the temperature $T_1$ and $T_a$, where the cuts correspond, with the same color scheme, to the curves plotted in panel (a) for $\chi=5$. \textbf{(d)} A color map of the differential conductance $\kappa_{\gamma}$ as defined in Eq. \ref{eq:g-gamma}, as a function of $T_1$ and $T_a$, where the NDTC region (where $\kappa_{\gamma} < 0$, in red) is visible; the cuts correspond, with the same color scheme, to the curves plotted in panel (b) for $\chi=5$.}
    \label{fig:1-ndtc}
\end{figure}

Fig. \ref{fig:1-ndtc} shows that the source-island channel in the device achieves NDTC. The photonic heat current $J_{\gamma}$ follows Eq. \ref{eq:Jgamma} and is plotted in Fig. \ref{fig:1-ndtc}(a,c) as a function of $T_1$, for different values of $T_a$, (keeping $T_s=\qty{1.4}{\kelvin}$, as in Fig. \ref{fig:0-intro}(b)). For each horizontal cut, NDTC is achieved for values of $T_1$ to the left of the maximum: In these intervals, increasing the temperature of the colder reservoir increases $J_{\gamma}$. At very low temperatures, $J_{\gamma}$ saturates to low values determined by the frequency-dependent part of $\rho$, while at high enough $T_1$ the DC part of $\rho$ dominates, as could already be deduced from Fig. \ref{fig:0-intro}(b,c).

We can also appreciate how the position and height of the maximum depend on $T_a$ from Fig. \ref{fig:1-ndtc}(c), where the horizontal cuts correspond to the plots of Fig. \ref{fig:1-ndtc}(a) with $\chi=5$. From this figure, we see that for higher values of $T_a$, this maximum becomes smaller but shifts toward higher values of $T_1$. This is because, for lower values of $T_a$, the transmission coefficient $\tau$ increases faster with $T_1$. For $T_a=\qty{50}{\kelvin}$ we also plot $J_{\gamma}$ with $\chi=1$ (continuous line) and $\chi=0.2$ (dotted line), showing how heat transport is less efficient, as expected from the corresponding values of $\tau$. In addition, the maximum height is affected: the greater $\chi$, the better the matching when $T_1$ is low, and the temperature gradient is high, so there is a greater total heat flux.
\color{black}

After we demonstrated the temperature dependence of the heat current, we can examine the differential thermal conductance of the photonic channel. It is defined as the derivative with respect to $T_1$ (with negative sign since $T_1$ is the lower temperature in the considered setup) of $J_{\gamma}$ (see Ref. \cite{casati_2006}):
\begin{eqnarray}
    \kappa_{\gamma} = -\frac{\partial J_{\gamma}}{\partial T_1}.
    \label{eq:g-gamma}
\end{eqnarray}
Indeed, $\kappa_{\gamma}$ may achieve negative values, as shown in Fig. \ref{fig:1-ndtc}(b), where it is plotted for $T_s=\qty{1.4}{\kelvin}$ while varying $T_1$. These negative values correspond to the NDTC range of $T_1$ described in Fig. \ref{fig:1-ndtc}(a,c). We can observe that the steeper $\tau$ becomes, the more negative the minimum of $\kappa_{\gamma}$. 
In mismatched cases, $\kappa_{\gamma}$ drops at negative values deeper and more rapidly in $T_1$ for values of $\chi>1$ and the opposite when $\chi<1$.

A more comprehensive view of the dependence of $\kappa_{\gamma}$ on $T_1$ and $T_a$ is given in Fig. \ref{fig:1-ndtc}(d). The figure represents the region of the $T_1$-$T_a$ plane in red, where the photon heat transport channel exhibits NDTC. This region is somewhat extended for every value of $T_a$, ensuring that the effect is robust with respect to the properties of the material. The position and value of the minimum of $\kappa_{\gamma}$ depend on the value of $T_a$. In particular, we note that the NDTC effect is more intense for lower values of $T_a$. 

Having acknowledged the effects of $\chi$ on the range and value of NDTC, we will always assume that it equals 5, a value compatible with an efficient thermal coupling between the reservoirs and the good performance of the proposed devices below.

\section{\label{sec:device}Device and temperature control}

\subsection{Thermal transistor}

We integrate the structure described in Sec. \ref{sec:ndtc} in a three-terminal device, like the one in Fig. \ref{fig:0-intro}(c), to obtain a thermal amplifier following the blueprint of Refs. \cite{casati_2006} and \cite{fornieri_2016}. We propose a device where the ``S" reservoir plays the role of the source terminal and ``1" is the central island, connected to the other two terminals, respectively, a gate ``G" and a drain ``D". The drain and the gate are metallic reservoirs with a tunnel connection to the central island, with tunnel resistances $R_d$ and $R_g$ and temperatures $T_d$ and $T_g$. We will assume that the drain reservoir is well thermalized to the substrate phonon bath (owing to a strong electron-phonon coupling or large volumes of the terminals), so we will always take $T_d=T_b$. We will consider, along with the photonic heat flow $J_{\gamma}$, the other heat currents associated with the central part of the device: the input gate flow $J_g$ from the gate, the output drain current $J_d$ in the drain at temperature $T_d$, and the heat $J_{ph}$ dissipated in the phonon bath. The arrows indicate the directions in which the respective currents are considered positive. 

We assume that the regions of the VRH reservoirs corresponding to the tunnel contacts are metallic due to heavy local doping, similar to the contacts in Sec. \ref{sec:ndtc}, allowing us to describe the energy exchange with the gate and drain using the equations for heat transport across a NIN junction, where the tunnel electrical resistance is temperature-independent \cite{giazotto_2006}. In this case, the expression for such tunnel heat current $J_i$ ($i = g,d$) between a hot terminal at temperature $T_{hot}$ and a cold one at $T_{cold}$ is:
\begin{eqnarray}
    \label{eq:Jt} J_{i}(R_t;T_{hot},T_{cold}) = \frac{\pi^2 k_B^2}{6 e^2 R_i}(T_{hot}^2 - T_{cold}^2),
    \label{eq:Jtunnel}
\end{eqnarray}
which is determined by the quasiparticle tunneling between two Fermi liquids kept at different temperatures, \textit{i.e.}, an NIN tunnel junction with electrical resistance $R_i$ \footnote{In those formulae, we are simply assuming that Wiedemann-Franz laws apply to the junction. However, our model requires only that the heat conductance is linear with the temperature gradient, \textit{i.e.}, Fourier laws.}.

To complete our description, we consider the heat lost in the phonon bath. $J_{ph}$ depends on the temperatures $T_1$ and $T_b$ in this way (supposing vanishing Kapitza resistance for the central island):
\begin{eqnarray}
    \label{eq:Jph} J_{ph}(\Sigma,V;T_1,T_b) = \Sigma V (T_1^6 - T_b^6),
    \end{eqnarray}
which is a model in line with Ref. \cite{beeman_1990, ventura_2006}, where $\Sigma$ is a parameter that describes the intensity of the electron-phonon thermal coupling, and $V$ is the volume of the central island, with $\Sigma \approx \qty{1e7}{\watt  \metre^{-3} \kelvin^{-6}}$.

In the PHA, the temperature of the central island, and consequently the heat transport properties, is determined by the balance of the incoming and outcoming heat currents, whose expressions are reported in Eqs. \ref{eq:Jgamma}, \ref{eq:Jt}-\ref{eq:Jph}. 
Once $T_s$, $T_g$, and $T_d = T_b$ are fixed, the equation:
\begin{eqnarray}
    \nonumber
    & J_{\gamma}(T_a;T_s,T_1) + & J_t(R_g;T_g,T_1) + \\
    & - J_t(R_d;T_1,T_d) & - J_{ph}(\Sigma,V;T_1,T_b) = 0,
    \label{eq:balance}
\end{eqnarray}
once solved numerically, determines the solution $T_1$.

\subsection{Parameters and temperature control}

We will focus on two possible parameter configurations, shown in Tab. \ref{tab:params}. These configurations are optimized to amplify gate current modulations (CMA) or gate temperature modulations (TMA). The choice of parameters was made to provide a configuration that offers both raw performance and an extended operating range, while being free from hysteresis effects.

\begin{table}
\begin{ruledtabular}
\begin{tabular}{l c c c c c c c c}
Device & $T_a$ (\unit{\kelvin}) & $T_s$ (\unit{\kelvin}) & $R_g$ (\unit{\kilo \ohm}) & $R_d$ (\unit{\kilo \ohm}) &  $V$ (\unit{\metre^3}) & $T_b$ (\unit{\kelvin}) & $\chi$ \\ 
 \hline
CMA & 50 & 1.4 & 40 & 20 & $10^{-18}$ & 0.02 & 5 \\
TMA & 50 & 1.4 & 20 & 20 & 5$\times 10^{-19}$ & 0.02 & 5
\end{tabular}
\end{ruledtabular}
\caption{\label{tab:params} The choices of parameters for the Current Modulation Amplifier (CMA) and the Temperature Modulation Amplifier (TMA). $T_a$ is the constant that appears in Eq. \ref{eq:ai-rho}; $T_s$ is the source temperature; $R_g$ and $R_d$ the tunnel resistances to the gate and drain terminals, respectively; $V$ the volume of the central island, $T_b$ the phonon bath temperature, and $\chi$ the geometric mismatch factor of Eq. \ref{eq:ai-tau}.}
\end{table}

\begin{figure*}[t!]
    \centering
    \includegraphics[width=\linewidth]{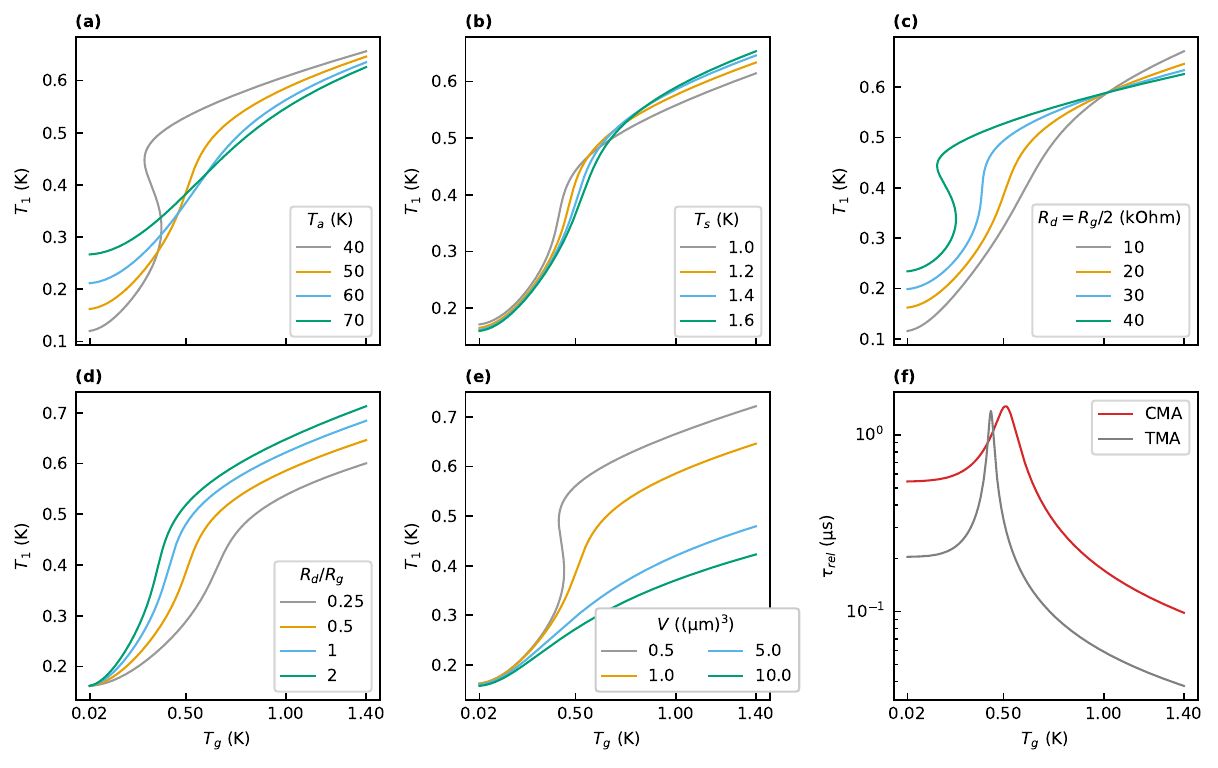}
    \caption{\textbf{(a)}-\textbf{(e)} The sets of points in the $T_1$, $T_g$ plane for which the balance of Eq. \ref{eq:balance} is satisfied for different choices of the device parameters (one per image, reported in the legend),  with the others kept as reported in the first line of Tab. \ref{tab:params} \textit{e.g.} $T_a=\qty{50}{\kelvin}$; in (d) the ratio of the resistances is varied while keeping the resistance of their parallel constant and equal to \qty{13.3}{\kilo \ohm}. \textbf{(f)} The dependence of the relaxation time conservative estimate $\tau_{rel}$ on the gate temperature $T_g$, for the CMA (red) and the TMA (gray), in logarithmic scale.}
    \label{fig:2-device}
\end{figure*}

To illustrate the influence of the parameters on performance, the solutions of Eq. \ref{eq:balance} in the $T_g$-$T_1$ plane are plotted in Fig. \ref{fig:2-device}(a-e) for various configurations. In particular, in each panel, one or more device parameters are varied while the others are taken from the first line of Tab. \ref{tab:params}. For now, we will not change $T_b=T_d$, to describe the performance of the device in a fixed range of gate temperatures (from $\qty{20}{\milli \kelvin}$ to $\qty{1.4}{\kelvin}$). However, we will see that the amplifying behavior is relatively robust with respect to the bath temperature in Secs. \ref{sec:j-ampli}-\ref{sec:t-ampli}.

In Fig. \ref{fig:2-device}(a), we present the behavior of $T_1$ as a function of $T_g$ for various values of $T_a$. 
As also in Fig. \ref{fig:2-device}(c,e), we can see that in some cases there are multiple possible values of $T_1$ for a single value of $T_g$. In this case, the PHA will show thermal hysteresis in $ T_1$. To avoid such a regime and obtain history-independent properties, we will prevent this regime, which is one reason behind the proposed devices' parameter choice. However, this hysteretic behavior may be utilized to define a thermal memory (a possibility we will discuss in future research). 
In addition, the choice of $T_a$ influences the minimum value of $T_1$ and the steepness of $T_1(T_g)$ (when it can be defined). A smaller $T_a$ means that $\rho$ as defined in Eq. \ref{eq:rho-omega} is more affected by $T_1$ at low temperatures, which in turn translates into a better mismatch between the source and the central island for $T_1 \approx \qty{20}{\milli \kelvin}$ and lower values for $T_1$. For analogous reasons, the curves become less steep as higher values of $T_a$ are chosen.

In Fig. \ref{fig:2-device}(a), we also show how changing $T_a$ affects the steepness and position of the inflection point in $T_1(T_g)$, which is pivotal for the range of operation of the device and its performances. In fact, around such a point, $ T_1$ is more responsive to variations in the temperature or current of the gate and influences the behavior of the entire device, achieving amplification. This steepness is also affected by the operating point of $T_s$ (see Fig. \ref{fig:2-device}(b)), the values of the tunnel resistances $R_g$ and $R_d$ (Fig. \ref{fig:2-device}(c)) and their ratio $R_d/R_g$ (Fig. \ref{fig:2-device}(d)).

The source temperature $T_s$ is the only parameter that can be varied while using the device at fixed bath and drain temperatures. As shown in Fig. \ref{fig:2-device}(b), it acts on both the inflection point's position and the temperature dependence's steepness. So, regulating $T_s$ is a way to tune the PHA.
This is similar to an electrical transistor, where the performance is partially tuned by changing the bias at the source terminal.

Tunnel resistance values are also relevant. In particular, it is possible to see (Fig. \ref{fig:2-device}(c)) that higher values of $R_d$ and $R_g$, which reduce the thermal contact of the island with the gate and the drain, favor a steep increase in $T_1$ as soon as $\tau$ is high enough. In comparison, lower values of tunnel resistance smooth such behavior. As seen in Fig. \ref{fig:2-device}(d), the ratio between the resistances also plays a role since the lower $R_d$, the better the coupling to the drain (which is the coldest terminal): thus, $T_1$ gets lower (when the ratio $R_d/R_g$ is varied, it is done by keeping the resistance of their parallel constant and equal to \qty{13.3}{\kilo \ohm} to make a more meaningful comparison).

Lastly, in Fig. \ref{fig:2-device}(e), the role of the central island volume $V$, which determines the coupling strength to the phonon bath (Eq. \ref{eq:Jph}), is highlighted. The smaller it is, the less coupled with phonons the central island becomes, resulting in a more significant excursion of $T_1$.

\subsection{\label{sec:relnoise}Amplifier bandwidth}

Another important figure of merit in the device is the relaxation time, $\tau_{rel}$, for the temperature of the electrons on the central island to assess the amplifier's bandwidth. This indicates the time required for the electronic temperature to achieve stationary equilibrium and respond to temperature variations at the terminals. An approximate linear estimation of this time can be given by \cite{vischi_2020}:
\begin{equation}
\tau_{rel} = \frac{C_1}{\kappa_{tot}},
\end{equation} 
where $C_1$ is the thermal capacitance of the electron of the central island, which is the product between the island volume $V$ and its specific heat $c_1 = \gamma T_1$, which is linear, despite the soft gap, since at low-energies the thermal excitation of electron-hole pairs in the disordered semiconductor dominates \cite{efros_1975, ventura_2006}. However, this specific heat could become flatter at lower temperatures due to the formation of spin-exchange clusters \cite{mccammon_2005}, but this is outside the operating range of our proposal. Plausible values are $\gamma=\qty{1}{\joule \m^{-3} \kelvin^{-2}}$~\cite{pesty_2000}. $\kappa_{tot}$ is the total thermal conductance between the electronic degrees of freedom of the reservoirs involved.

A conservative first estimate for $\tau_{rel}$ can be given by accounting only for the contribution of the phonon heat channel, approximating $\kappa_{tot}\approx\kappa_{ph} = \partial J_{ph} / \partial T_1$. So we get $\tau_{rel,ph} = \gamma / 6 \Sigma T_1^4$, which is volume independent ($\kappa_{ph}$ is evaluated in Appendix \ref{app:conds}). 
Taking $T_1 = \qty{20}{\milli \kelvin}$, we get the maximum $\tau_{ph} \approx \qty{100}{\milli \second}$. Knowing that phonon heat transport is relatively inefficient at low temperatures, a more precise evaluation can be given by estimating $\kappa_{tot}$ as the sum of the following conductances:
\begin{eqnarray}
    \kappa_{tot} = & \kappa_d + \kappa_{\gamma} + \kappa_g + \kappa_{ph} = \nonumber \\
    = & \frac{\partial J_{d}}{\partial T_1} - \frac{\partial J_{\gamma}}{\partial T_1} + \frac{\partial J_{g}}{\partial T_1} + \frac{\partial J_{ph}}{\partial T_1},
\end{eqnarray}
which are calculated in Eq. \ref{eq:g-gamma} and Appendix \ref{app:conds}. 
The temperature dependence of $\kappa_{tot}$ determines a non-monotonic behavior of the relaxation time $\tau_{rel}$ with the temperature, being
\begin{eqnarray}
    \tau_{rel} = \frac{\gamma V T_1}{\kappa_{tot}(T_1)},
\end{eqnarray}
which is shown in Fig. \ref{fig:2-device}(f), where we plotted $\tau_{rel}$ for the two different  device proposals. This non-monotonic behavior directly reflects that
$\kappa_{\gamma}$, starting from low temperatures, arrives at a minimum negative value as presented in Fig. \ref{fig:1-ndtc}. Instead, all other thermal conductances $\kappa_d$, $\kappa_g$, and $\kappa_{ph}$, respectively, toward the drain, the gate, and the phonon baths, increase with $T_1$. 
When $\kappa_{\gamma} < 0$, a positive temperature fluctuation $\Delta T_1 > 0$ increases $J_{\gamma}$. 
The increase in heat flux from the source partially compensates for the rebalancing losses, 
slowing down the relaxation. 

The relaxation time determines the device's bandwidth since $\Delta f \approx 1/\tau_{rel}$. Even in the worst case, the result is of the order of \unit{\mega \hertz}, putting the PHA on par with the molecular thermal switches already described in the literature \cite{li_2023}. At the same time, it also offers the functionalities of an amplifier.

\section{\label{sec:th-ampli}Thermal amplifiers}

Thanks to the knowledge gained from exploring the parameter space in the previous section, it is possible to propose devices that achieve optimal performance. In the following, we will first study the CMA device in Sec. \ref{sec:j-ampli}. After that, leveraging that a slight change in $T_g$ can strongly alter $T_1$, we propose the TMA device in Sec. \ref{sec:t-ampli}. Lastly, we show an alternative, fully photonic setup in Sec. \ref{sec:fullphoton}.

\subsection{\label{sec:j-ampli}Current Modulation Amplifier (CMA)}

The CMA, thanks to an apt choice of parameters (see Tab. \ref{tab:params} for its specifications), responds to a slight oscillation on the gate thermal current $J_g$, inducing a variation on $J_{\gamma}$ and $J_d$, possibly with an amplification ratio.  
So, we can introduce the amplification factors ($i =\gamma,d$)
\begin{equation}
\alpha_{i}=\bigg|\frac{\partial J_{i}}{\partial J_g}\bigg|,
\end{equation} 
which can be computed in a linear approximation as~\cite{fornieri_2016}, using the heat current conservation $J_g=J_d-J_{\gamma}+J_{ph}$:
    \begin{eqnarray}
    \label{eq:alpha} 
    \alpha_i = \bigg| \frac{\partial J_i}{\partial T_1} \frac{\partial T_1}{\partial (J_d - J_{\gamma} +J_{ph})}\bigg| = \bigg| \frac{\kappa_i}{\kappa_d + \kappa_{\gamma} + \kappa_{ph}} \bigg|, 
    \end{eqnarray}
where $i=\gamma,d$.

Eq. \ref{eq:alpha} clarifies why the NDTC regime allows us to obtain a heat current amplifier $\alpha_i>1$. This requires that the numerators are more significant than the denominators. However, as long as $\kappa_{\gamma}$ and $\kappa_d$ are positive, this will not be the case, but if $\kappa_{\gamma}<0$, this becomes possible~\cite{casati_2006}.

\begin{figure}[t!]
    \centering
    \includegraphics[width=\linewidth]{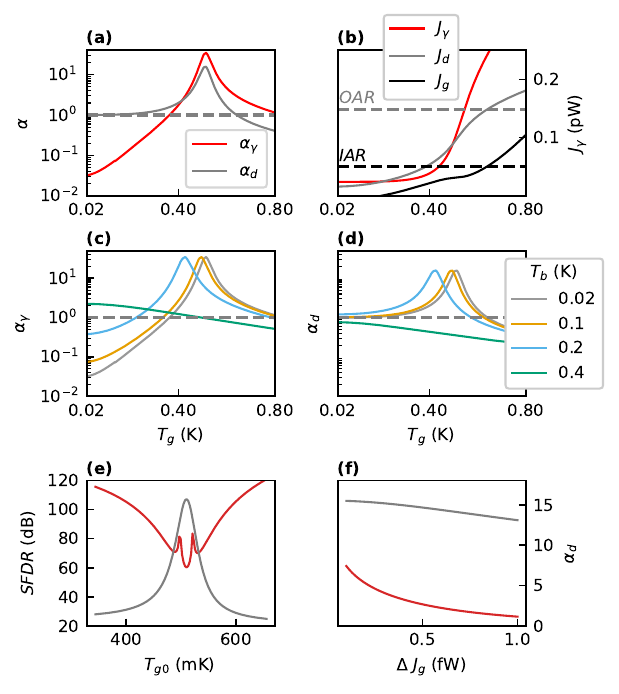}
    \caption{The plots of this image represent the behavior of the CMA with the parameters of the first line of Tab. \ref{tab:params}; \textbf{(a)} The amplification factors $\alpha_{\gamma}$ and $\alpha_d$ as a function of $T_g$ for a bath temperature $T_b=\qty{20}{\milli \kelvin}$ in logarithmic scale, with the gray dashed line $\alpha=1$ that highlights the region where the amplification factors are greater than 1. \textbf{(b)} The values of $J_{\gamma}$, $J_d$ and $J_g$, in the same configuration of panel (a), with the maximum gate current of the $IAR_{CMA}$ and the maximum drain current of the $OAR_{CMA}$ indicated respectively by the black and the gray dashed lines. \textbf{(c)}-\textbf{(d)} $\alpha_{\gamma}$ and $\alpha_d$ at various bath temperatures $T_b$. \textbf{(e)} The $SFDR$ (red) and the gain $\alpha_d$ (gray) for an input heat current $J_g$ modulation of amplitude $\qty{0.1}{\femto \watt}$ as a function of the working point $T_{g0}$ at $T_b = \qty{20}{\milli \kelvin}$.  \textbf{(f)} The $SFDR$ (red) and the gain $\alpha_d$ (gray) at the point of maximum gain $T_{g0}=\qty{510}{\milli \kelvin}$ as a function of the amplitude $\Delta J_g$ of the input current $J_g$ modulation at $T_b = \qty{20}{\milli \kelvin}$.}
    \label{fig:3-jampli}
\end{figure}

With our specific choice of parameters, it is possible to achieve amplification factors as great as 15 at both the source and the drain, as seen in Fig. \ref{fig:3-jampli}(a). The maximum gain value is achieved at $T_g=\qty{510}{\milli \kelvin}$.

In analogy to Ref. \cite{paolucci_2017}, to evaluate the performance of the device, we introduce the figure of merit of the ``Input Active Range" $IAR_{CMA}$. This is the range of $J_g$ values for which the amplification factor at the drain exceeds 1. The $IAR_{CMA}$ goes from 0 to the black dashed line in Fig. \ref{fig:3-jampli}(b). At $T_b=\qty{20}{\milli \kelvin}$, $IAR_{CMA}= \qty{50}{\femto \watt}$. This interval is mainly limited because the of $\tau$ saturating, hindering the sensitivity to $T_1$ and the NDTC effect.

We can also evaluate the ``Output Active Range" $OAR_{CMA}$ as the range of the possible $J_d$ values that the PHA can feed as output while in the amplifying regime. At $T_b=\qty{20}{\milli \kelvin}$, $OAR_{CMA}=\qty{150}{\femto \watt}$. This range of drain current values goes from 0 to the gray dashed line in Fig. \ref{fig:3-jampli}(b). We also notice that, while we define and evaluate these merit figures referring to the drain current, the gate input shows a similar degree of control over the source current, apart from being less effective for low values of $J_g$ and $T_g$.

It is, in principle, possible to obtain higher amplification factors by adequately adjusting the CMA's parameters, even arriving at diverging values in Eq. \ref{eq:alpha}. Higher gains are obtained when a greater responsivity of $T_1$ is achieved to variations of $T_g$: that is, for the choices of parameters that make the curves of Fig. \ref{fig:2-device}(a-e) steeper. 
However, divergent gains come with a price, as they strongly distort the amplifier's linearity. So, we show a typical configuration in which the gain is still substantial, but less signal distortion is introduced. Those parameters also achieve better values of the input range and avoid saturation.

In Fig. \ref{fig:3-jampli}(b), the values of the currents at play are represented. In particular, it is possible to see that a smaller gate current $J_g$ drives higher fluxes in the amplifying regime. It is also interesting to estimate the switching ratio $\eta$ \cite{li_2023_trans} between the ``OFF" and ``ON" state of the PHA ($i = \gamma,d$)
\begin{eqnarray}
    \eta_{i} = \frac{J_{i}^{\textrm{ON}}}{J_{i}^{\textrm{OFF}}}.
\end{eqnarray}
To evaluate such figure of merit in a configuration where $J_g$ is lower than $J_{\gamma,d}$, we propose $T_g = T_b = \qty{20}{\milli \kelvin}$ as the OFF state and $T_g = \qty{640}{\milli \kelvin}$ (the upper limit of the amplification regime) as the ON state. For both source and drain currents, $\eta_i \sim 10$. This could be increased with designs and parameters tailored for efficient switching, not amplification.

In Fig. \ref{fig:3-jampli}(c-d), $\alpha_{\gamma}(T_g)$ and $\alpha_d(T_g)$ are plotted, using the parameters of Tab. \ref{tab:params}, except for $T_b$, which is varied to show the resilience of PHA amplifier to the background temperature setting. In particular, the amplifying behavior near the maximum remains unchanged up to $T_b=\qty{100}{\milli \kelvin}$. Even at $T_b = \qty{200}{\milli \kelvin}$, the theoretical maximum value of the amplification factors is close to 15, but it would require an unusual configuration with the gate cooler than the bath. 

In the PHA, amplification of an input modulation may happen with distortion. To estimate this, we evaluate the ``Spurious-Free Dynamic Range", which is related to the ratio between the first harmonic $A_1$ and the second largest harmonic $A_M$ of the device's response $J_d$ to a small sinusoidal modulation of the input $J_g$. According to the formula in Ref. \cite{desimoni_2023}:
\begin{eqnarray}
\label{SFDR}
    SFDR = 20 \log_{10} \bigg(\frac{A_1}{A_M}\bigg),
\end{eqnarray}
which is expressed in \unit{\deci \bel}.
More details on the estimate of this merit figure are given in the Appendix \ref{app:sfdr}.
In Fig. \ref{fig:3-jampli}(e-f) the $SFDR$ (in red) is plotted, together with the drain gain $\alpha_d$ (in gray), as a function of the working point $T_{g0}$, for a small modulation of the input current of amplitude $\qty{0.1}{\femto \watt}$ (Fig. \ref{fig:3-jampli}(e)) and as a function of the modulation of the amplitude $\Delta J_g$ at the point of maximum gain $T_{g0}=\qty{510}{\milli \kelvin}$ (Fig. \ref{fig:3-jampli}(f)). We observe that the point of maximum gain corresponds with the greatest anharmonicity and that the linearity of the response decreases with the modulation amplitude. However, we notice that the $SFDR > \qty{20}{\deci \bel}$ and $\alpha_d > 10$ up to $\Delta J_g=\qty{1}{\femto \watt}$, which corresponds to a modulation of the 3\% on the total input current $J_{g0}(T_{g0}=\qty{510}{\milli \kelvin})\approx \qty{30}{\femto \watt}$.

With the modalities presented in Appendix \ref{app:noise}, we also estimate the CMA's Noise Equivalent Power (NEP). In particular, in this case, for the work point $T_g=\qty{510}{\milli \kelvin}$, the input NEP is $\textrm{NEP}_{CMA}=\qty{2.0e-7}{\pico \watt \per \sqrt{\hertz}}$.

It should be noted that the amplifying behavior is not dependent on the fact that the source and the central island are in this specific variable range hopping regime; it can also be obtained for a Mott insulator (where the fraction $1/T$ in the exponent increases to a power of 1/4 \cite{efros_1984}), although with the drawbacks of less control over the amplification factors and a more limited input range. 

\subsection{\label{sec:t-ampli}Temperature Modulation Amplifier (TMA)}

In this specific configuration, a slight variation of $T_g$ can induce a more remarkable change in $T_1$.
In particular, in this section, we present how the TMA device (with the parameters of Tab. \ref{tab:params}) achieves this. 

\begin{figure}[t!]
    \centering
    \includegraphics[width=\linewidth]{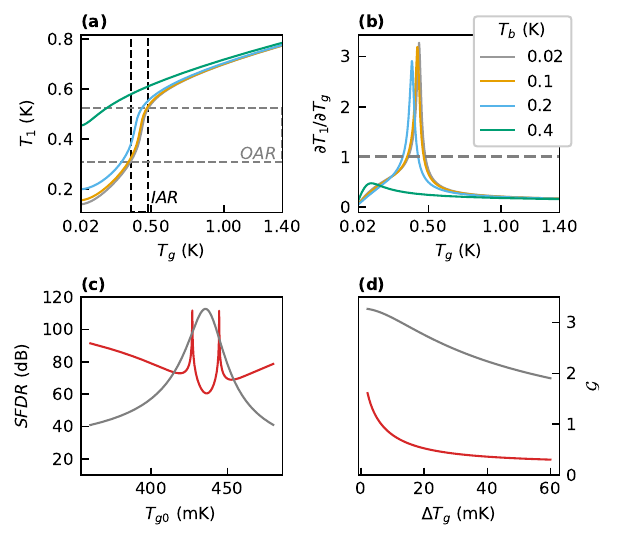}
    \caption{The plots of this image represent the behavior of a device with the parameters of the second line of Tab. \ref{tab:params}; \textbf{(a)} The dependence of $T_1$ on $T_g$, for various bath temperatures $T_b$, with the $T_g$ range of the $IAR_{TMA}$ at $T_b=\qty{20}{\milli \kelvin}$ delimited by black dashed lines and the corresponding $T_1$s of the $OAR_{TMA}$ by gray dashed lines. \textbf{(b)} the differential gain $\partial T_1/ \partial T_g$ as a function of $T_g$, with a gray dashed line dividing the region of actual gain where $\partial T_1/ \partial T_g>1$ from the region of lower gain, also in this case plotted for different values of $T_b$. \textbf{(c)} The $SFDR$ (red) and the gain $\mathcal{G}$ (gray) for an input modulation of $T_g$ of amplitude $\qty{2}{\milli \kelvin}$ as a function of the working point $T_{g0}$ at $T_b = \qty{20}{\milli \kelvin}$. \textbf{(d)} The $SFDR$ (red) and the gain $\mathcal{G}$ (gray) at the point of maximum gain $T_{g0}=\qty{435}{\milli \kelvin}$ as a function of the amplitude $\Delta T_g$ of the input temperature modulation on $T_g$ at $T_b = \qty{20}{\milli \kelvin}$.}
    \label{fig:4-Tampli}
\end{figure}

In such a device, the fundamental figure of merit is the differential gain $\mathcal{G}$ that, in the linear approximation, is the ratio between the output $T_1$, and the input $T_g$ oscillation amplitudes, \textit{i.e.}
\begin{eqnarray}
    \mathcal{G} = \frac{\partial T_1}{\partial T_g},
\end{eqnarray}
keeping fixed all the other temperatures $(T_s,T_d,T_b)$. For the device we are considering, $T_1(T_g)$ (Fig. \ref{fig:4-Tampli}(a)) and the corresponding $\partial T_1/ \partial T_g$ (Fig. \ref{fig:4-Tampli}(b)) are shown for a selection of bath temperatures $T_b$. Up to $T_b=\qty{100}{\milli \kelvin}$ the amplitude of the output oscillations can be more than triple the input ones, with a maximum value $\mathcal{G}(T_g = \qty{435}{\milli \kelvin})=3.3$ achieved for the coldest bath.

We can define  $IAR_{TMA}$ and  $OAR_{TMA}$ in this case, in analogy to what is done in Sec. \ref{sec:j-ampli}, with the difference that they are expressed in temperature ranges. At $T_b = T_d = \qty{20}{\milli \kelvin}$ the $IAR_{TMA} = \qty{120}{\milli \kelvin}$ (calculated as the width of the interval of $T_g$s for which $\mathcal{G}>1$, delimited by black dashed lines in Fig. \ref{fig:4-Tampli}(a)), higher with respect to other examples of temperature amplifiers \cite{paolucci_2017}, where the $IAR$ is of the order of tens of \unit{\milli \kelvin}. Referring to the $T_1$ outputs corresponding to this input range, $OAR_{TMA} = \qty{215}{\milli \kelvin}$, as can be measured from the data of Fig. \ref{fig:4-Tampli}(a), where gray dashed lines delimit them.

In Fig. \ref{fig:4-Tampli}(c-d),  $SFDR$ (red lines) is plotted alongside an estimate of $\mathcal{G}$ (gray lines), given as the ratio between the first harmonic of the output $A_1$ and the amplitude $\Delta T_g$ of the input. In this case, as the Appendix \ref{app:sfdr} explains, we analyze the response regarding the central island temperature $T_1$ to a sinusoidal modulation of $T_g$ at a specific working point. These quantities are plotted in two ways: as a function of the average gate temperature $T_{g0}$ and the modulation amplitude $\Delta T_g$. When the modulation amplitude is considered, we took $T_{g0}=\qty{435}{\milli \kelvin}$, which is the point of maximum gain and correspondingly should also be the point of greatest anharmonicity, as anticipated in Fig. \ref{fig:4-Tampli}(c). In particular, it is interesting to see that even for modulations of tens of \unit{\milli \kelvin}, the anharmonicity is kept under control, 
$\gtrsim \qty{20}{\deci \bel}$ between the carrier and the second greatest harmonic. At the same time, the gain remains above 2.

Also, in this case, as shown in the Appendix \ref{app:noise}, it is possible to estimate the total noise spectral density on the central island and infer the NEP on the input at the gate. In particular, dividing by the thermal conductance, we can assume the noise equivalent temperature oscillation $\textrm{NET}_{TMA} \approx \qty{6.7}{\micro \kelvin \per \sqrt{\hertz}}$. This NET aligns with the sensitivities of advanced solid-state thermometers \cite{gasparinetti_2015} and qualifies the TMA as a promising preamplification stage for bolometers.

\subsection{\label{sec:fullphoton}Fully photonic design}

\begin{figure}[t!]
    \centering
    \includegraphics[width=0.8\linewidth]{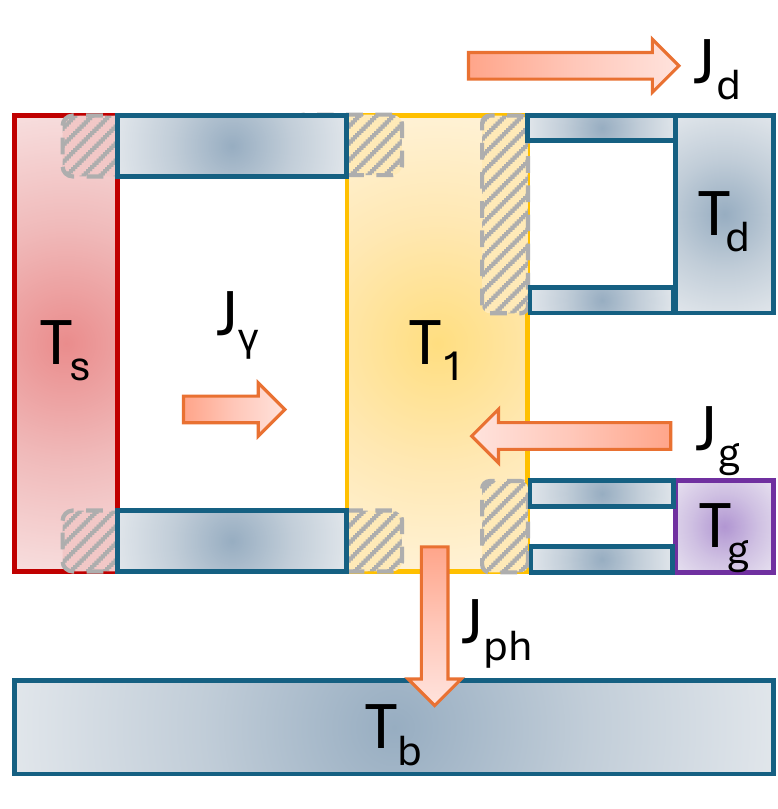}
    \caption{A possible variation on the design of Fig. \ref{fig:0-intro} features tunnel barriers replaced by lossless lines that are thermalized to the phonon bath, closing the loop on an Ohmic region of the reservoir. This design enhances the flexibility of photonic heat transport and its long-range connectivity to the drain and gate channels.}
    \label{fig:5-fullphoton}
\end{figure}

The blueprint we provided above allows for numerous developments. One of the most promising possibilities is a device in which photons constitute all heat transport channels between the terminals. This design variation is illustrated in Fig. \ref{fig:5-fullphoton}.
Such a device can operate thanks to the possibility of having heavily doped, locally metallic regions within the material, as described in Sec. \ref{sec:ndtc}. Suppose that we connect a metallic reservoir, such as the gate or the drain, to a single, continuous metallic region in the central island through lossless lines. In that case, we create a photonic heat transport channel between two Ohmic impedances whose transmission coefficient is nearly unaffected by temperature variations. This differs from the source-island connection because, in that scenario, the Ohmic regions in contact with the lines are separated and in series with the bulk of the reservoir. 

The degree of freedom provided by the choice of $R_g$ and $R_d$ is replaced by the matching of the impedances of the Ohmic regions and the corresponding terminals, which impacts the $\tau$ of the specific channel according to Eq. \ref{eq:tau}. 
This proposal follows the same temperature scaling for all heat channels as the previously proposed devices so that we can expect similar performances.
A fully photonic PHA can connect reservoirs over great distances, utilizing the advantages of photon-mediated heat transport \cite{partanen_2016} while offering a flexible design that allows tunable elements to control transmission. In principle, this geometry has a limited flow of heat (fixed by the quantum of heat conductance) in the photonic channel. However, this is also true for the previous proposal, at least for the source current, where that heat channel is still photonic.
Furthermore,  in the setup of Fig. \ref{fig:5-fullphoton}, capacitors can be eventually integrated into the lines, providing a full non-galvanic coupling to the terminals. This opens the possibility to electrically isolate in DC the different terminals but still operating as a thermal amplifier, while mitigating cross-talk and ground loop issues.

\section{\label{sec:concl}Conclusions}

This paper proposes a device that achieves NDTC using electromagnetic fluctuations for heat exchange. Although it has been observed and utilized in a heat amplifier based on near-field radiation \cite{benabdallah_2014}, a mesoscale integrated solid state demonstration of NDTC and heat amplification is still lacking. Our proposal is a possible answer to this gap. 
Depending on the configuration, the PHA can function as a heat current modulation or a temperature modulation amplifier. We show that this device can achieve amplification factors up to $15$ in the former regime and $3$ in the latter. The device exhibits low relaxation times, indicating that its operational speed aligns with other microscopic thermal devices tested experimentally \cite{li_2023}.

The PHA may represent one of the essential components for executing a fully thermal logic, as previously suggested within phononic \cite{wang_2007} and phase-coherent electronic \cite{paolucci_2018} heat transport frameworks, with electromagnetic fluctuations propagating in lossless wires. Relying on these fluctuations also means that they are compatible with the established framework of circuit quantum electrodynamics \cite{wallraff2020cqed} and thermodynamics \cite{pekola2014cqtd}. This architecture can manage and harvest the waste heat from other devices or the attenuation stages of electrical lines \cite{li_2023_trans} to power other processes. Complete control of spurious heat can be helpful in power operations that have already been demonstrated, such as refrigerators compatible with photon-mediated heat transport \cite{gasparinetti2025autonomous, mottonen2024autonomous}.

Autonomous temperature control based on fully thermal logic and feedback can be envisioned. The PHA's performance as a thermal transistor, which allows control of the source-drain current by the gate temperature, can also be exploited for automatic on-chip heat regulation. 
For example, the device could switch to the ``ON" state when the gate temperature exceeds a certain threshold and cool the source terminal by letting a significant amount of heat flow out of it. 

Furthermore, in its temperature modulation amplifier configuration, the PHA presents an intriguing possibility for sensing and monitoring device temperatures, serving as an amplification stage for on-chip thermometers. For instance, in the case of bolometers, such amplification enables the detection of more minor temperature fluctuations without affecting the heat capacity of the absorber, thereby enhancing its sensitivity. The performance of the device are compatible with the state-of-the-art, including new protocols for a broad-range bolometer-enabled superconducting circuit spectroscopy \cite{pekola2025readout}. Due to its capability of mapping an oscillation in the IAR to the OAR, which is a different temperature range, the TMA could also be used to translate thermal signals to the point of maximum sensitivity of a sensor.

Compared with other mesoscopic devices, our proposal is based on photonic transport along lossless lines, which operates efficiently at the cryogenic temperatures needed for specific implementations of quantum technologies and can also be integrated into solid-state setups. Our design can be adapted either to harness electron- and photon-mediated heat transport or to entirely rely on the latter, as discussed in Sec. \ref{sec:fullphoton}. 

Studies have shown that heat transport along superconducting lines remains effective over long distances and can be controlled by incorporating tunable impedances \cite{partanen_2016,ronzani2018tunable,subero2023resistive,maillet_2020}. 
The use of photons as heat carriers also offers the advantage of avoiding galvanic coupling to target devices \cite{giazotto_2017_capacitive}, a feature that could be beneficial when high coherence is essential, such as in qubit thermal management \cite{yoshioka_2023, nakamura2025fastqcr}. 
Another potential direction is to integrate multiple parallel lossless lines into the design to enhance the conductivity between the source and the central island, thus improving power performance.

Other applications of this design can take advantage of its bistability, achievable by carefully selecting parameters, especially $T_a$. The two possible equilibrium states in this configuration can eventually function as the states of a register \cite{kubytskyi_2014}. 

\begin{acknowledgments}
We acknowledge the EU’s Horizon 2020 Research and Innovation Framework Programme under Grants No. 964398 (SUPERGATE), No. 101057977 (SPECTRUM), and the PNRR MUR project PE0000023-NQSTI for partial financial support. A.B. acknowledges the MUR-PRIN2022 Project NEThEQS (Grant No. 2022B9P8LN) and CNR project QTHERMONANO. 
\end{acknowledgments}

\appendix

\section{\label{app:conds}Thermal conductances}

The expressions of the thermal conductances of the central island with respect to all other reservoirs employed in the main text are obtained by deriving the thermal current formulas of $T_1$ for the different heat transport channels. We also correctly considered the current conventions adopted in the main text (see Sec. \ref{sec:ndtc}). 

The $\kappa_{\gamma}$, fundamental in the functioning of the PHA, is already discussed in the main text of Eq. \ref{eq:g-gamma} due to its relevance. 
The heat conductances of the tunnel contacts with the gate and the drain, respectively $\kappa_g$ and $\kappa_d$, are obtained by deriving Eq. \ref{eq:Jt}:
\begin{eqnarray}
    \kappa_{g,d} = \bigg| \frac{\partial J_{g,d}}{\partial T_1} \bigg| = \frac{\pi^2 k_B^2}{3 e^2 R_{g,d}}T_1.
\end{eqnarray}
Given the adopted heat current convention, we took the absolute value because we wanted to compute the differential flow of heat outside of the central island.
Instead, the heat conductance lost in the phononic degrees of freedom is: 
\begin{eqnarray}
    \kappa_{ph} = \frac{\partial J_{ph}}{\partial T_1} = 6 \Sigma V T_1^5,
\end{eqnarray}
simply obtained by the derivative of Eq. \ref{eq:Jph}. The phononic contribution decreases faster than the other conductances at low temperatures, becoming less important as the temperature decreases.

\section{\label{app:sfdr}$SFDR$ estimate}

The $SFDR$ is computed, as told in Secs. \ref{sec:j-ampli} - \ref{sec:t-ampli}, from the coefficients of the Fourier series of the responses to a sinusoidal input. In particular, for the CMA, the expression of the input is the following: 
\begin{eqnarray}
    J_{g,in,CMA}(T_{g0},\Delta J_g;t) & = & J_{g0}(T_{g0}) + \Delta J_g \sin(t),
\end{eqnarray}
where $J_{g0}$ is the gate current in the system when $T_g=T_{g0}$. 

Instead, the TMA input is written as:
\begin{eqnarray}
    T_{g,in,TMA}(T_{g0},\Delta T_g;t) & = & T_{g0} + \Delta T_g \sin(t).
\end{eqnarray} 

Then, the respective outputs, $J_d(t)$ for the CMA and $T_1(t)$ for the TMA, can be evaluated by numerically solving Eq. \ref{eq:balance} for $T_1$ and, in the case of $J_d(t)$, substituting the result back into the expression of the drain current.
The numerically computed Fourier series coefficients of these outputs $A_i$, where $i$ is a multiple of the input frequency, are used to calculate $SFDR$ according to the methodology discussed in Eq.~\ref{SFDR}.

\section{\label{app:noise}Noise estimate}
We suppose that, for frequencies below the \unit{\mega \hertz}, which is the bandwidth of the device, as seen in Sec. \ref{sec:relnoise}, the electron gas of the VRH islands has time to thermalize and relax the hot spots that may form inside the material. In such a limit, for sufficiently low temperatures $T_i\ll T_a$, the noise spectral density follows the standard expressions \cite{rogovin1974fluct,golubev_2001}. We calculated the low-frequency power fluctuations spectral densities from the tunnel contacts and the phonon bath to the central island by following Refs. \cite{golubev_2001} and \cite{vischi_2020}. The expression is computed by using the fact that a tunnel junction follows Poissonian statistics, so noise density is given ($i=g,d$) by:
    \begin{eqnarray}
     \nonumber \mathcal{N}^2_{i} = \frac{2}{e^2 R_{i}} \int_{-\infty}^{+\infty} \epsilon^2 [f(\epsilon,T_{i}) + \\
     + f(\epsilon, T_1) - 2f(\epsilon,T_{i})f(\epsilon, T_1)] d\epsilon,
\end{eqnarray}
where $f(\epsilon,T)$ is the Fermi distribution at energy $\epsilon$ and temperature $T$. It is important to note that because of the tunneling nature of the problem, the same general formula applies to the case of terminals with strong (nonlinear) temperature differences.

Instead, the spectral thermal noise density due to the thermal contact with the phonon bath is \cite{vischi_2020}:
\begin{eqnarray}
     \mathcal{N}^2_{ph} & = & 12 k_B V \Sigma (T_b^7 + T_1^7),
\end{eqnarray}
where we have considered the appropriate power laws of the electron-phonon coupling of NTD germanium \cite{beeman_1990}. 

The investigation of noise and fluctuation in systems with substantial temperature differences has only recently been extensively discussed \cite{golubev_2015,Lumbroso_2018,Sivre_2019,Larocque_2020,tesser_2023,Pierattelli_2024}. However, for the power fluctuations of the photonic channel, we refer
to the framework recently presented in Ref. \cite{basko_2024}. As it can be seen in Fig. \ref{fig:0-intro}(c), for the operational range of our devices, the frequency dependence is a minor correction, so, for simplicity, we will assume a frequency-independent model of two purely dissipative impedances (pure resistances). Indeed, for tiny fluctuations, we obtained an expression for the spectral density of the fluctuations on the central island due to the source terminal heat inflow
\begin{widetext}
    \begin{eqnarray}
     \mathcal{N}^2_{\gamma} = \frac{\tau}{4 \pi \hbar} \int_{-\infty}^{+\infty} \epsilon^2 \bigg[ \coth\bigg(\frac{\epsilon}{2 k_B T_s}\bigg)\coth\bigg(\frac{\epsilon}{2 k_B T_1}\bigg) -1 \bigg] d\epsilon,
    \end{eqnarray}
\end{widetext}
where $\tau$ is taken outside of the integral as we are approximating it as frequency (energy) independent.
Note that all these expressions, for identical temperatures $T$, correspond to the thermal noise limit $\mathcal{N}^2_i = 4 k_B \kappa_i T^2$, where the subscript $i = \gamma, g, d, ph$ identifies the heat transport channel of conductance $\kappa_i$ \cite{giazotto_2006}.

The noise estimate of Sec. \ref{sec:j-ampli} is obtained from the components of the power fluctuations on the central island presented above, evaluated for the CMA device. The results are given for a gate temperature of \qty{510}{\milli \kelvin}, for which the device achieves the maximum amplification factor of around 15 for the drain and source outputs.
We obtained, respectively, for the gate, the drain, the phonon bath, and the source contributions, the following values:
\begin{eqnarray}
   \mathcal{N}^2_{g,CMA} \approx \qty{3.3e-12}{(\pico \watt)^2 \per \hertz},\\
   \mathcal{N}^2_{d,CMA} \approx \qty{2.3e-12}{(\pico \watt)^2 \per \hertz},\\
   \mathcal{N}^2_{ph,CMA} \approx \qty{2.6e-12}{(\pico \watt)^2 \per  \hertz}, \\
   \mathcal{N}^2_{\gamma,CMA} \approx \qty{6.0e-13}{(\pico \watt)^2 \per \hertz}.
\end{eqnarray}
These values, assumed to be uncorrelated, sum up to a total of a fluctuation spectral density on the output of $\mathcal{N}^2_{CMA,out} \approx \qty{8.8e-12}{(\pico \watt)^2 \per \hertz}$. Considering an amplification factor of 15, the Noise Equivalent Power on the input results in being 
\[\textrm{NEP}_{CMA} = \sqrt{\mathcal{N}^2_{CMA,out}}/15 \approx \qty{2.0e-7}{\pico \watt \per \sqrt{\hertz}}.\]

In the same way, it is possible to evaluate the contributions of the low-frequency fluctuation spectral density of the various components in the TMA device (Sec. \ref{sec:t-ampli}). In particular, estimating the noise-equivalent temperature fluctuation at the TMA input (gate) is possible. In this case, we evaluate the four contributions for the configuration of maximum gain at $T_g=\qty{435}{\milli \kelvin}$, we get
 \begin{eqnarray}
   \mathcal{N}^2_{g,TMA} \approx \qty{5.6e-12}{(\pico \watt)^2 \per \hertz},\\
   \mathcal{N}^2_{d,TMA} \approx \qty{3.1e-12}{(\pico \watt)^2 \per \hertz},\\
   \mathcal{N}^2_{ph,TMA} \approx \qty{8.4e-13}{(\pico \watt)^2 \per \hertz},\\
   \mathcal{N}^2_{\gamma,TMA} \approx \qty{1.1e-12}{(\pico \watt)^2 \per \hertz}.
\end{eqnarray}
We then take the square root of their sum (since we again assume independent fluctuations) and convert it to the output $T_1$ temperature oscillations by dividing by the total thermal differential conductance \cite{richards_1994} $\kappa_{tot} = \qty{0.16}{\pico \watt / \kelvin}$ (evaluated as the sum of the conductance formulas presented in Appendix \ref{app:conds}) so that we get the amplitude of the corresponding temperature oscillations on the output \cite{gasparinetti_2015}, which we can then be divided by the gain of $\mathcal{G} \approx 3$ to obtain the input equivalent spectral density 
\[\mathrm{NET}_{TMA} \approx \qty{6.7}{\micro \kelvin \per \sqrt{\hertz}},\]
\,\\
which is compatible with the expected noise performances of the most advanced low-temperature solid-state thermometers \cite{gasparinetti_2015}.

\end{document}